\documentclass[twocolumn]{aastex631}
\usepackage{amsmath}
\usepackage{amssymb}

\usepackage{ wasysym }
\usepackage{bm}
\usepackage[dvipsnames]{xcolor}

\begin{document}
 
\title{Intermediate-Mass Black Hole Formation from Hierarchical Mergers in Galactic Nuclei}

\correspondingauthor{Amanda Newton}
\email{anewton@luc.edu}

\author[0009-0001-1211-2072]{Amanda Newton}
\affil{Loyola University Chicago, 6460 N Kenmore Ave, Chicago, IL 60660, USA}
\affiliation{Center for Interdisciplinary Exploration and Research in Astrophysics (CIERA) and Department of Physics and Astronomy, Northwestern University, 2145 Sheridan Road, Evanston, IL 60201, USA }

\author[0000-0003-0984-4456]{Sanaea C. Rose}
\affiliation{Center for Interdisciplinary Exploration and Research in Astrophysics (CIERA) and Department of Physics and Astronomy, Northwestern University, 2145 Sheridan Road, Evanston, IL 60201, USA }

\author[0000-0003-4412-2176]{Fulya K{\i}ro\u{g}lu}
\affiliation{Center for Interdisciplinary Exploration and Research in Astrophysics (CIERA) and Department of Physics and Astronomy, Northwestern University, 2145 Sheridan Road, Evanston, IL 60201, USA }

\author[0000-0003-0992-0033]{Bao Minh Hoang}
\affiliation{Department of Physics and Astronomy, University of California, Los Angeles, Los Angeles, CA 90095, USA}
\affil{Kall Morris Inc, 130 W Washington St, Marquette, MI 49855, USA}

\author{Frederic A.\ Rasio}
\affiliation{Center for Interdisciplinary Exploration and Research in Astrophysics (CIERA) and Department of Physics and Astronomy, Northwestern University, 2145 Sheridan Road, Evanston, IL 60201, USA }

\begin{abstract}

Dense stellar environments like nuclear star clusters (NSCs) can dynamically assemble gravitational wave (GW) sources. We consider a population of single stellar mass black holes (BHs) in the inner $0.1$~pc of a NSC surrounding a $4 \times 10^6$~M$_\odot$ supermassive black hole (SMBH). Using a semianalytic model, we account for direct collisions between BHs and stars and GW capture between BHs. We explore the effect of the initial BH mass and spin distributions on their final properties and the production of GW sources. Specifically, we consider upper and lower limits for the BH initial mass distribution, and we account for the possibility that a subset of our initial population are the merger products of primordial BH binaries. We find that $\sim 500$ M$_{\odot}$ intermediate mass black holes (IMBHs) can form for our upper limit mass distribution, while our lower limit mass distribution forms none. Most IMBHs $\gtrsim 200$~M$_\odot$ eventually sink towards the center of the cluster and merge with the SMBH. We also find successive BH-star collisions can produce low-spinning BHs with $\chi \lesssim 0.2$.  Our results have implications for LIGO-Virgo-KAGRA sources. We find that the overall merger rate depends primarily on the initial BH mass distribution and is $\gtrsim 10^{-9}$~yr$^{-1}$ per Milky Way-like galaxy for our range of initial conditions. However, primordial binaries can change the number of GW mergers with second and higher generation progenitor BHs by an order of magnitude.


\end{abstract}

\keywords{Galactic center (565), Gravitational wave sources (677), Intermediate mass black holes (816), Stellar mass black holes (1611), Stellar dynamics (1596)}

\section{Introduction} \label{sec:intro}


The detection of gravitational waves (GWs) from merging compact objects has fundamentally changed the way we sense the Universe. Detections by the LIGO-Virgo-KAGRA Collaboration have provided new insights into the demographics of black holes (BHs) \citep[e.g.,][]{Fishbach+20}, even in some cases challenging our previous understanding of these populations. Event GW190521, for example, produced an intermediate-mass BH (IMBH) \citep{LIGO2}, as did the recent event GW231123, the most massive merger detected to date \citep{LIGO_mostmassive_2025}. Furthermore, the progenitor BHs in these events lie in the pair-instability mass gap, above the maximum BH mass predicted by many stellar evolution models \citep[e.g.,][]{Heger2003,Spera+17,Woosley2017,Limongi+18,Belczynski+20,Renzo+20,Sakstein+20,Vink+21}, while several GW events, including GW231123, have components with non-zero spins, indicating dynamical formation \citep[e.g.][]{LIGO_mostmassive_2025,Tong+25_dynamicalformation}.  


Dense stellar clusters have long been recognized as important contributors to GW sources \citep[e.g.,][]{quinlan1987,Rodriguez+18,Rodriguez+19}. The detected highly spinning BHs, several of which lie in the pair-instability mass gap, may be the product of sequential BH mergers from close dynamical interactions in dense clusters  \citep[e.g.,][]{AntoniniRasio2016,Fishbac+17,Fragione+20,Kimball+21_evidenceforhierarchicalmergers}. However, GW recoil kicks can pose a problem for globular clusters with typical escape speeds of 50 km/s; the velocity kick from asymmetric GW emission can eject merger products from the cluster \citep[e.g.,][]{OLeary+06,Schnittman+07,Centrella+10,Baibhav+20}. Recently, \citet{Mai+25} showed that the most massive globular clusters can retain merger products and go on to form second generation and higher mergers similar to those observed by LIGO-Virgo-KAGRA. Similarly, nuclear star clusters (NSCs) have deep potential wells and can retain merger products, as has been shown for NSCs without supermassive black holes (SMBHs) \citep[e.g.,][]{MillerLauburg09,AntoniniRasio2016,Fragione+21}. 
In one relevant dynamical process, single BHs which approach each other closely may radiate enough energy via GWs to become a bound system. \citet{OLeary+09} find GW capture rates of $\sim 10^{-10}$ to $\sim 10^{-9}$ yr $^{-1}$ per Milky Way-like galaxy for a population of single BHs in a NSC with a SMBH,
though they did not detail the specific merger generation or mass and spin properties of those events. More recently, \citet{liu2025} considered a broad range of dynamical interactions, focusing on binary-single interactions that form second generation/second generation (2G/2G) mergers in NSCs.

The presence of a SMBH can increase the potential for these environments to generate GW sources. A SMBH can induce binary systems in its vicinity to merge by exciting their orbital eccentricities, in a process known as the Eccentric Kozai-Lidov (EKL) Mechanism \citep[e.g.,][]{Kozai,Lidov,Naoz+13sec,Naoz+13GR,Naoz16}. This process can lead to the merger of BH binaries which, if left on their initial orbit, would otherwise not have merged within a Hubble time \citep[e.g.,][]{AntoniniPerets12,Antognini14,Hoang+18,Fragione+18,Grishin+25,Stegmann+25,Yubo2025_spinsKozai}. In addition to BH-BH interactions, dense stellar environments facilitate direct collisions between BHs and stars \citep[e.g.,][]{BaileyDavies99,Davies+11,Rose+22,Kiroglu+25_BHstarspins}. \citet{Rose+22} show that frequent collisions between BHs and stars in nuclear star clusters can create more massive BHs, some $> 100$~M$_\odot$ \citep[for a similar process in globular clusters, see][]{Rizzuto+22}. In addition to shaping the BH mass demographics, the subsequent accretion of material from the star can have detectable implications for the BH spins \citep{Lopez2019,Kiroglu+25_highspinsaccretion,Kiroglu+25_spinorbit}. 

Future and ongoing GW observational campaigns motivate our study of the rates and properties of merger events from NSCs, as well as the processes shaping the broader BH population in these environments. We expand the model in \citet{Rose+22}, which accounts for BH-star collisions and two-body relaxation, to include GW capture between single BHs. 
We vary the initial BH mass and spin distributions and consider the possibility that a fraction of the BHs in our initial population come from already-merged primordial binaries, based on the results of \citet{Hoang+18}. Our goal is to determine how the initial conditions affect the dynamical fate of the BHs, including the production of GW sources, and map onto the final property distributions. We also assess the feasibility of forming IMBHs (BHs with mass $>100$ M$_{\odot}$) in NSCs. Our paper is organized as follows:

In Section~\ref{sec: semianalytic_model}, we give an overview of our semianalytic model for the evolution of BHs in a NSC with an SMBH. In particular, Section \ref{sec:initialconditions} describes our initial conditions and Section~\ref{sec:processes} describes the physical processes we consider in a cluster. We present and discuss results in Section \ref{sec:results}, including signatures of dynamical processes in the mass and spin distributions of the BHs in Sections~\ref{sec:BHstar_coll_spin_effects} and \ref{sec:mass_and_spin_distributions}, IMBH formation in Section~\ref{sec:IMBHformation}, and implications for GW sources in Section~\ref{sec:GW_rates_and_properties}.

\section{Overview of Semianalytic Model}
\label{sec: semianalytic_model}

We use a semianalytic model first developed by \citet{Rose+22} to study the dynamical evolution of a cluster of BHs in a NSC like the Milky Way's, with a central $4 \times 10^6 M_{\astrosun}$ SMBH. For the purposes of this study, we assume the NSC is composed of 1 $M_{\astrosun}$ stars and BHs with initial masses described in Section~\ref{sec:initialconditions}. We follow a sample of $1000$ BHs embedded in a NSC over an integration time of $10$ billion years, tracking their properties at each timestep. We draw their initial masses, spins, and orbital properties statistically as described in Section~\ref{sec:initialconditions} so that they are representative of the overall BH population.

While we only follow a sample of $1000$ BHs, our model accounts for the presence and effects of the surrounding star cluster using a statistical approach and timescale arguments. The surrounding cluster can be described by two key properties, the density and velocity dispersion, which vary with distance from the SMBH.
We describe the velocity dispersion of objects using:

\begin{equation}
    \sigma = \sqrt{\frac{GM_{\bullet}}{(\alpha+1)r}}
    \label{eq:velocity_disp}
\end{equation}
where $M_\bullet$ is the mass of the SMBH and $r$ is the distance from the SMBH. The velocity dispersion also weakly depends on $\alpha$, the slope of the density profile, described below.

The number density of the BHs depends on the interplay of different dynamical processes and interactions with other objects. Generally, the number density profile of objects can be described by a power law, $n \propto r^{-\alpha}$. For a single-mass population, the objects will eventually settle onto an equilibrium distribution with $\alpha = 1.75$ \citep{BahcallWolf76}. In a population with a spectrum of masses, the heavier objects tend to sink towards the SMBH due to energy equipartition. We adopt density profiles for a two-component population, BHs and relatively lighter stars, inspired by work in the literature \citep{AharonPerets16,LinialSari22}. The number density of the BHs depends on distance from the SMBH. We use the BH density profile calculated using a Fokker-Plank model in \citet{AharonPerets16}, shown in their Figure 1 left panel. The density can be described a by power law:

\begin{equation} \label{eq:BH_density_power_law}
n_\text{BH} = n_0 \left(\frac{r}{R_\text{h}} \right)^{-\alpha},
\end{equation}

where $n_0$ is a normalization constant 1.1$\times$10$^{-2}$pc$^{-3}$,
$R_\text{h}$ is the sphere of influence of the cluster, 1 pc, and $\alpha$ is the slope of the density profile, chosen to be $\alpha \approx 1.83$, based on results from \citet{AharonPerets16}.

While they consider a four component population (main-sequence stars, BHs, white dwarfs, and neutron stars), their BH cusp is comparable to the analytic result for a two-component population from \citet{Rom+2024} in the inner $0.1$~pc region of the NSC.







We note that they do differ outside of this region, but as the dynamical processes operate on long ($>10$~Gyr) timescales outside of $0.1$~pc, it will not significantly change our results. A comparison of these two density profiles is presented in Appendix \ref{app:n_BH_comp}, showing that the power law from \citet{AharonPerets16} works as a sufficient approximation of the \citet{Rom+2024} density within the inner 0.1 pc. The profile yields $\gtrsim 1000$ BHs in the inner $0.1$~pc of the NSC, consistent with our choice to follow $1000$ BHs in this region. 

Most models predict that the relatively lighter stars should lie on a shallower cusp with index $\alpha \lesssim 1.5$ \citep[eg.][]{BahcallWolf76, AlexanderHopman+09,Preto+10,LinialSari22}. However, collisions between the stars themselves can deplete the stars \citep{Mastrobuono-Battisti+21, Rose+23, Sidhu+2025,kim2026}, and observations of the Milky Way's nuclear star cluster have indicated a range of $\alpha$ between roughly $1.1$ and $1.4$ \citep{Habibi+19, Schodel+20}.
We use a power law similar to Eq. \eqref{eq:BH_density_power_law} for stellar density:
\begin{equation} \label{eq:density_power_law}
\rho = \rho_0 \left(\frac{r}{r_0} \right)^{-\alpha}
\end{equation}
where $r$ is the distance from the SMBH in pc, $r_0$ is a normalization value in pc, and $\alpha$ is the density slope. We adopt $r_0 = 0.25$ pc and $\rho_0 = 1.35 \times 10^6 M_{\odot} \, \, \text{pc}^{-3}$ to normalize the power-law \citep{Genzel2010} and $\alpha = 1.25$ for the slope of the profile. 
Our rational for the value of $\alpha$ is that this choice can only lead to a lower star-BH collision rate due to a lower stellar density, and therefore a more conservative estimate of BH growth through this channel. 


\section{Black Hole Initial Properties and Binarity} \label{sec:initialconditions}

\begin{figure*}
    \centering
    \includegraphics[width=0.75\linewidth]{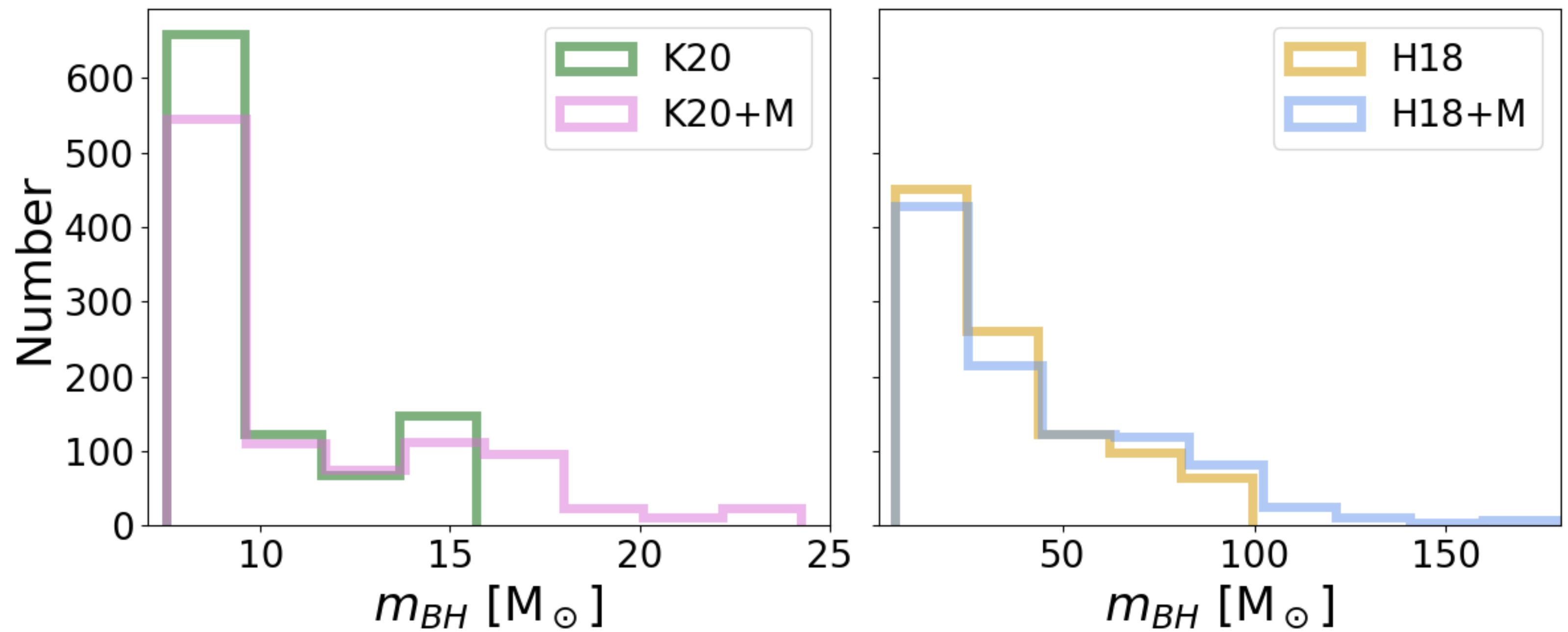}
    \caption{We show the distributions for the four initial BH mass distributions described in Section \ref{sec: semianalytic_model}. \textbf{Left:} The conservative case of all initially low mass stellar BHs from \cite{Kremer+20,Belczynski+16} (K20 ), and an additional case where 15\% of the single BHs are products of a primordial binary merger (K20 \raisebox{0.3ex}{\footnotesize{$+$}}M). \textbf{Right:} The upper mass limit case from \citet{Hoang+18}. One case includes primordial binaries and single BHs (H18\raisebox{0.3ex}{\footnotesize{$+$}}M), and H18 describes a case where BHs are evaporated and initially single.}   \label{fig:IC_plot}
\end{figure*}

We use sets of initial conditions based on \citet{Hoang+18}. While we treat a population of single BHs around the SMBH, it is plausible that some of the BHs initially resided in binaries. We distinguish these binaries as primordial systems, meaning they are the end products of stellar binary evolution as opposed to dynamical formation. Any binary system in the central parsec of the Galactic center will experience coherent gravitational perturbations from the SMBH. The SMBH can torque the binary's orbit
and induce binary mergers \citep[for a review of the EKL mechanism, see][]{Naoz16}. The combination of this effect and mergers from GW emission alone, based on the initial binary configuration, can lead to roughly $15 \%$ of BH binaries merging within the NSC \citep{Hoang+18}. Binaries in the NSC are also subject to frequent weak gravitational interactions with other surrounding stars, and in a process called binary evaporation, these interactions can unbind some of the binaries \citep[e.g.,][]{Heggie75,Stephan+16,Rose+20}. \citet{Hoang+18} study a population of BH binaries within the sphere of influence of the SMBH self-consistently. We therefore account for the presence of primordial binaries as 15\% of the BH population, using their final conditions as our initial conditions.

We consider two cases for BH initial mass and spin distributions. In the first case, we use the same initial mass distribution as \citet{Hoang+18}, but assume all the BHs were single from birth and have zero initial spin. We refer to this case as H18. In the second case, we compute the final masses and spins (see Section \ref{sec:GWcap_methodology} for equations) of the merger products from \citet{Hoang+18} and combine them with their single BHs from the unbound primordial binaries. We refer to this case as H18\raisebox{0.3ex}{\footnotesize{${+}$}}M. The overwhelming majority of their primordial binaries either merge or become unbound within $<1$ billion years, making the single BH population a reasonable starting condition for our $\sim 10$ billion year timescale simulation. We note that they also have a subset of primordial binaries that neither merge nor become unbound. These systems constitute $\lesssim 40 \%$ of their initial systems, and at the end of their simulations, single BHs outnumbered binaries by about $2.5$ to $1$, meaning the population of BHs at later times should be dominated by singles. As dynamical processes related specifically to binary systems, such as binary-single interactions, are beyond the scope of this study, we reserve the inclusion of these systems for future work. 


\citet{Hoang+18}'s initial mass distribution extends up to $\sim 90$ M$_\odot$. In addition to the first two sets of initial mass and spin distributions described above, we also consider a more conservative lower limit BH mass distribution based on Cluster Monte Carlo (CMC) globular cluster simulations by \citet{Kremer+20} \citep[see also,][]{Belczynski+16}. We assume solar metalliticy for all stars. 
Similar to above, we have one set of initial conditions where all BHs are single and non-spinning and another set where a subset of the BHs are the products of primordial mergers.
Specifically, we take our lower limit mass distribution and randomly pair and merge $15 \%$ of the BHs so that the same fraction have had a previous primordial BH-BH merger, either EKL-induced or through GW emission alone, as in the simulations of \citet{Hoang+18}. To determine final masses and spins of these remnants, we use the prescription for binary BH mergers described in Section \ref{sec:GWcap_methodology}. While GW merger rate can depend on BH mass \citep[see][]{Hoang+18}, we hold the fraction of merged systems fixed for a direct comparison of initial conditions. In particular, a third of the primordial binary systems merged in this population, as well as in H18\raisebox{0.3ex}{\footnotesize{$+$}}M. We refer to these cases as K20  and K20 \raisebox{0.3ex}{\footnotesize{$+$}}M, respectively.

These four distributions are shown as histograms in Figure \ref{fig:IC_plot}. We see that H18\raisebox{0.3ex}{\footnotesize{$+$}}M allows for BH mass as high as 175 $M_{\odot}$, while K20 \raisebox{0.3ex}{\footnotesize{$+$}}M only extends to $25$~$M_{\odot}$. While we do not show the initial spin distributions in Figure~\ref{fig:IC_plot}, the initial conditions with mergers have a peak around 0.7, consistent with \citep[e.g.,][]{Fishbac+17, Antonini+2019, GerosaFishbach2021, Tagawa+2021, Borchers+2025, Mai+25}. 
The initial spin distributions can be seen in Figure~\ref{fig:final_mass_and_spins_dist}, where we discuss their evolution over the course of the simulations.

\section{Modeling Physical Processes within the Cluster}
\label{sec:processes}

We account for a number of dynamical processes in tandem within the cluster using a statistical approach. Each physical process in the cluster has an associated timescale that depends on the stellar density profile, $\alpha$, and velocity dispersion, $\sigma$. We describe our treatment of the stellar dynamics below.

\subsection{Gravitational Wave Capture Between Single Black Holes}
\label{sec:GWcap_methodology}

BH-BH binaries form from GW emission from close encounters between single BHs. During these close encounters, the BHs must radiate enough energy away through GWs to become a bound system
\citep[e.g.,][]{quinlan1987,OLeary+09}. 
The maximum impact parameter to produce a merger is $b_{\text{max}}$:
\begin{equation}
b_{\text{max}} = \left(\frac{340\pi\eta}{3}\right)^{1/7}\frac{GM_{\text{tot}}}{c^2}\left(\frac{v_{\text{rel}}}{c}\right)^{-9/7},
\label{eq:merger_bmax}
\end{equation}
where $M_{\text{tot}}$ is the total mass of the two compact objects, $\eta$ is their symmetric mass ratio, $\eta = m_1 m_2/(m_1 + m_2)^2$, and $v_{\text{rel}}$ is relative velocity at infinity, which we take to be the velocity dispersion given by Eq.~\eqref{eq:velocity_disp} \citep{OLeary+09,Gondan+18a,Hoang+20}. We note that there is also a minimum impact parameter, $b_{\text{min}}$,
\begin{equation}
b_{\text{min}} = \frac{4GM_{\text{tot}}}{c^2}\left(\frac{v_{\text{rel}}}{c}\right)^{-1},
\label{eq:merger_bmin}
\end{equation}
for which any interaction with $b < b_{\text{min}}$ will be a direct collision.
Therefore, the GW capture cross section is:
\begin{equation}
A_{\rm cap} = \pi(b_{\rm max}^2 - b_{\rm min}^2)
\label{eq:merger_GW_cross_sec}
\end{equation}
\citep[e.g.,][]{Hoang+20}. While direct collisions between BHs still result in a merger, the cross-section as set by $b_{\rm min}$ is so small that including or excluding these interactions does not change our results. With the cross-section for interaction, the timescale for GW capture can be calculated as:
\begin{equation}
t_{\rm GW} =  \left(A_{\rm cap} n_{\rm BH} \sigma \right)^{-1}, 
\label{eq:t_GW}
\end{equation}
where $n_{\rm BH}$ is the number density of BHs.


For each BH in our sample population, we estimate the timescale for GW capture from Eq.~\eqref{eq:t_GW} by evaluating $n_{\rm BH}$ and $\sigma$ at the BH's semimajor axis about the SMBH. We take $m_2$ in $\eta$ to be the average of the initial mass distribution. The probability of GW capture occurring over an interval of time $\Delta t$ is $\Delta t$/ $t_{\rm GW}$.  For choice of $\Delta t$, the integration timestep, see Section \ref{sec:stellcoll_methodology}. 
After determining the GW capture probability, we generate a random number between $0$ and $1$. If the number is less than or equal to the capture probability, we assume a GW capture has occurred \citep[emulating the methodology in][]{Rose+22}. Once we have determined statistically that GW capture has occurred, we draw a mass and spin for the second BH from the initial distribution. Calculations from \citet{OLeary+09} indicate that BH binaries formed through GW capture in NSCs form with high eccentricity and merge rapidly, before the BH binary can be dynamically affected by a passing object. Therefore, we assume the merger is prompt, within hours of the binary's formation. We then calculate the final mass, spin, and recoil kick using the same equations as the Cluster Monte Carlo Code \citep[e.g.,][]{Rodriguez+22}, which are drawn from numerical relativity and other studies \citep[e.g.,][]{Barausse&Rezzolla2009, Barausse+2012}.
Specifically, final merger remnant mass is determined by: 
\begin{multline}
\frac{m_{f}}{M_{\rm tot}} =
1 - \eta (1 - 4\eta)(1 - E_{\text{ISCO}}) \\
- 16 \eta^2 \left(p_0 + 4 p_1 \chi_\parallel (\chi_\parallel + 1) \right)
\label{eq:merger_mass_frac}
\end{multline}
where $E_{\rm ISCO} = \sqrt{1-2/3r_{\rm ISCO}}$ and the fitting constants are $p_0=0.04827$ and $p_1=0.01707$ \citep{Reisswig+2009, Barausse+2012,Rodriguez+18}. ISCO refers to the innermost stable circular orbit of matter accreting into the BH. $\chi_\parallel$ is the parallel component of the initial spin parameter, $\chi$. This parameter is normalized by $J/M^2$, where $J$ is the angular momentum and $M$ is the mass of the system \citep{Li2014}. $\chi_\parallel$ is defined as:
\begin{equation}
\chi_{\parallel} = \frac{m_1^2 \chi_1 + m_2^2 \chi_2}{(m_1 + m_2)^2} L
\label{eq:chipar_i}
\end{equation}
for BHs with masses $m_1$ and $m_2$ and spins $\chi_1$ and $\chi_2$.
Given $L$, the orbital angular momentum, we also calculate the final spin as:
\begin{equation}
\begin{aligned}
\chi_{f} = \min\Biggl( 1,\;
\Biggl| &
\frac{q^2 \chi_2 \cos\theta_2 + \chi_1 \cos\theta_1}{(1+q)^2} \\
& + \frac{q L}{(1+q)^2}
\Biggr|
\Biggr)
\end{aligned}
\label{eq:merger_af}
\end{equation}
where $q$  is the mass ratio of the BHs, $\cos{\theta_1} = \chi_1 L$, and $\cos{\theta_2} = \chi_2 L$.

$r_{\rm ISCO}$ is defined as: \

\begin{equation}
r_{\rm ISCO} = 3 + Z_2 \mp \sqrt{(3 - Z_1)(3 + Z_1 + 2Z_2)},
\label{eq:r_isco}
\end{equation}
where $Z_1$ and $Z_2$ are parameters dependent on $\chi$:
\begin{equation}
\begin{gathered}
Z_1 = 1 + (1-\chi^2)^{1/3} [(1+\chi)^{1/3} + (1-\chi)^{1/3}]
\\
Z_2 = [3\chi^2 + Z_1^2]^{1/2}.
\label{eq:spin_Z1_Z2}
\end{gathered}
\end{equation}





\subsubsection{Recoil kicks}

Depending on asymmetries in the system, GW radiation can be preferentially beamed in a particular direction, imparting a recoil kick on the merger product. We account for the dynamical effects of this recoil kick as follows. First, we begin by computing the components of velocity kick using equations from e.g., \citet{Holley-Bockelmann+08} 
\begin{equation}
    v_{\rm kick} = (1+e) [x(v_{\rm m}+v_{\perp} \cos\xi) + y v_{\perp} \sin\xi + zv_{\parallel}],
    \label{vkick}
\end{equation}
where
\begin{equation}
    v_{\rm m} = A \frac{q^2(1-q)}{(1+q)^5} \left(1+ B \frac{q}{(1+q)^2}\right),
    \label{vm}
\end{equation}
\begin{equation}
    v_\perp = H \frac{q^2}{(1+q)^5} (\chi_{\parallel,2} - q \chi_{\parallel,1}),
    \label{vperp}
\end{equation}
\begin{equation}
    v_\parallel = K \cos(\Theta - \Theta_0) \frac{q^2}{(1+q)^5} (\chi_{\perp,2} - q \chi_{\perp,1}),
    \label{vpara}
\end{equation}
and 
$\Theta$ refers to the angle of the direction of the merger, $\Theta_0$ refers to the angle of the initial direction of motion, and $\xi$ refers to the angle resulting from the unequal mass and spin contributions in the recoil kick. For each merger, these angles are chosen according to a uniform distribution, and the fitting constants, $A, B, H,$ and $K$ can be found in \citet{Holley-Bockelmann+08} \citep[see also, e.g.,][]{Campanelli+07, Gonzalez+2007,Lousto+08, Lousto+12, Gerosa+16}.
%
%
We apply the kick instantaneously at some point along the binary black hole's orbit about the SMBH. We draw the location of the merger statistically, weighted by the time spent at each point along the orbit. 
We calculate the new orbit of the BH merger product about the SMBH from the new mass and velocity vector of the merger product \citep[similar to applications of various types of kicks in][]{Naoz+22,Hoang+22,Jurado+24,RoseMockler2025}.

\subsection{Direct Collisions with Stars}
\label{sec:stellcoll_methodology}
Direct BH-star collisions can occur in NSCs. Each BH is expected to collide with stars over a characteristic timescale, the collision timescale:
\begin{equation}
t_{\rm coll} = \frac{1}{n_{\star}A\sigma},
\label{t_coll}
\end{equation}
where $n_{\star}$ is the number density of stars, $A$ is the cross-section of interaction,
and $\sigma$ as the velocity dispersion. The cross-section of interaction, $A$, is the geometric cross-section plus a term for gravitational focusing. The collision timescale becomes:
\begin{eqnarray} \label{eq:t_coll_full}
     t_{\rm coll}^{-1} &=& \pi n \sigma \nonumber \\ &\times& \left(f_1(e_{\rm BH})r_c^2 + f_2(e_{\rm BH})r_c \frac{2G(m_{\rm BH}+ M_\odot)}{\sigma^2}\right)\,
\end{eqnarray}
where $G$ is the gravitational constant and $r_c$ is the sum of the radii of the interacting objects, a black hole with mass $m_{\rm BH}$ and a star with mass 1 M$_\odot$.
As described in \citet{Rose+20}, $f_1(e_{\rm BH})$ and $f_2(e_{\rm BH})$ account for the effect of the BH's orbital eccentricity $e_{\rm BH}$ about the SMBH on the collision timescale, while $n$ and $\sigma$ are calculated from Eqs. \eqref{eq:density_power_law} and \eqref{eq:velocity_disp} at the semimajor axis of the orbit. We assume the stars in the NSC are a uniform population of $1$ $M_{\odot}$ stars such that $n_{\star} = \frac{\rho}{M_{\odot}}$ where $\rho$ is the stellar density profile, described by a power law (see Eq.~\eqref{eq:density_power_law}). The probability of a collision occurring over the integration timestep is $\Delta t$/ $t_{\rm coll}$. We initially set $\Delta t$ to be $10^6$ years \citep[for discussion, see][]{Rose+22}, however we adjust the timestep during the simulation to be always less than the collision timescale. 
Similarly to Section \ref{sec:GWcap_methodology} with GW capture, after the determining the collision probability, we generate a random number between $0$ and $1$. If the number is less than or equal to the collision probability, we assume a collision has occurred.

\subsubsection{Mass Growth through Collisions}
BHs may accrete mass during a direct collision with a star. Following each collision, we adjust the mass of the BH following \citet{Rose+22}. First, we estimate the amount of mass captured by the BH from the star based on Bondi-Hoyle accretion:
\begin{equation}
\dot{m}_{\rm BH} = \frac{4 \pi G^2 m_{ i}^2 \rho_\star}{(c_s^2+\sigma^2)^{3/2}}.
\label{eq:mass_change}
\end{equation}
where $m_{i}$ is the BH's initial mass, $\rho_\star$ is the star's density, and $c_s$ is the speed of sound within the star
\citep[g.g.,][]{BondiHoyle,Bondi52,Shima85,BondiHoyleOverview}.
We approximate $\rho_{\star}$ as $\frac{3 M_\odot}{4 \pi R_\odot^3}$ and take $c_s = 600~km~s^{-1}$ as the sound speed \citep{Christensen-Dalsgaard+96}. The amount of mass that the BH manages to capture from the star can then be approximated as:
\begin{equation} \label{eq:deltam}
    m_{\rm cap} = {\rm min}(\Delta m_{\rm BH}\times t_{\rm \star, cross~},~1~{\rm M}_\odot) \ ,
\end{equation}
where $t_{\rm \star, cross} \sim R_\star/\sigma$ is the crossing time of the BH within the star.
While the BH can capture mass from the star through direct collisions, radiative feedback can prevent it from accreting all of the stellar material. We estimate the accreted mass to be a factor $v_{\rm esc}/(c \eta)$ less than the captured mass, where $v_{\rm esc}$ is the escape velocity from the BH at $1$~R$_\odot$ and $\eta = 0.1$ is the accretion efficiency at the ISCO (see \citet{Rose+22} for more details and discussion of the accretion). We then adjust the BHs mass by $\Delta m_{\rm BH} = m_{\rm cap} \times v_{\rm esc}/(c \eta)$. 

\subsubsection{Spin Change through Collisions}
\label{sec:stellcoll_spin_methodology}
The spin of a BH also changes following a collision with a star due to angular momentum conservation during the accretion process. 
We use equations from \citet{Volonteri2013} to calculate the final spin of the BH, represented by $\chi$. The final spin $\chi_f$ of the BH is:
\begin{flalign*}
\hspace{5pt}\chi_f &= &
\end{flalign*}
\begin{equation}
\begin{cases} 
{\scriptstyle \frac{r_{\mathrm{ISCO}}^{1/3}}{3} \cdot \frac{m_{i}}{m_{f}} \left( 4 - \sqrt{\left( \frac{3 m_{i}^2 r}{m_{f}^2} \right) - 2} \right),} & \text{if } \frac{m_{f}}{m_{i}} \leq r^{1/2}, \\[8pt]
1, & \text{if } \frac{m_{f}}{m_{i}} \geq r^{1/2}.
\end{cases}
\label{eq:coll_af}
\end{equation}
where $\frac{m_{f}}{m_{i}}$ is the percent change in mass of the BH, and $r_{\rm ISCO}$ is defined in Eq. \eqref{eq:r_isco}.



\subsection{Relaxation and Dynamical Friction}

The orbital energy of a BH in a NSC changes over time through weak gravitational interactions with other objects. This process acts over the associated two-body relaxation timescale, which describes the time for the orbital energy and angular momentum to change by order of themselves: 
\begin{equation}
t_{\mathrm{relax}} = 0.34 \frac{\sigma^3}{G^2 \rho \langle M_{\rm avg}\rangle \ln\Lambda}. 
\label{eq:t_rlx}
\end{equation} $\ln\Lambda$ is the Coulomb logarithm, $\langle M_{\rm avg} \rangle$ is the average object mass, and $\rho$ is their mass density \citep[e.g.,][]{BinneyTremaine}. Occasionally, weak gravitational interactions cause BHs to wander too close to the SMBH, which eventually leads to inspiral into the SMBH and GW emission, an event known as an Extreme Mass Ratio Inspiral (EMRI)  \citep{MagorrianTremaine99, WangMerritt04, Hopman+05, AharonPerets16, StoneMetzger16, Amaro-Seoane18, SariFragione19, Naoz+22}. We model the relaxation process by simulating a random walk in the semimajor axis and eccentricity parameter space \citep[for the full equations, see][]{Naoz+22,Rose+22}.

GW recoil kicks can also alter the orbit of the BH about the SMBH. Depending on the strength of the kick and the BH's initial orbit, the merger product can be ejected from the inner $0.1$~pc of the cluster. If the merger product's new orbit remains bound to the SMBH, we allow it to sink back into the inner $0.1$~pc over a mass-segregation timescale: 
\begin{equation}
t_\mathrm{seg} \approx \frac{M_{\star}}{m_{\mathrm{BH}}} \times t_{\mathrm{relax}}(\langle M_{\rm avg}\rangle=M_{\star},~\rho = \rho_{\star})
\label{eq:t_seg}
\end{equation}
\citep{Spitzer1987, Fregeau+02, Merritt06}. This treatment of dynamical friction is consistent with previous semianalytic models of successive BH-BH mergers in NSCs \citep[as has been done in, e.g.,][]{AntoniniRasio2016,Fragione+21}. If the merger product is ejected from the cluster, it triggers a stopping condition in our code and we record the final mass and orbital configuration.



\subsection{Gravitational Wave Inspiral into the SMBH}

We also account for GW dissipation between each cluster BH and the SMBH. We calculate changes to the orbital eccentricity and semimajor axis due to GW emission over each timestep using equations from \citet{Peters+63,Peters64}. These changes become particularly important $<0.001$ pc of the SMBH, where the inspiral timescale is roughly less than simulation time of 10 billion years. This result is shown in \citet{Rom+2024} Figure 10.

Furthermore, we implement two stopping conditions which, if triggered, flag a BH as having undergone an EMRI. The first condition is triggered if the periapsis of the BH’s orbit about the SMBH falls within the critical radius $R_{\text{crit}} = 8Gm_{\text{SMBH}}/c^2$. The other condition classifies the BH as an EMRI if its remaining GW merger time with the SMBH is less than 100 years.

We note that our stopping conditions may classify BHs that are on plunging or eccentric orbits as EMRIs. However, we have verified that the majority of BHs we classify as EMRIs have already circularized when they trigger our stopping conditions. Excluding eccentric orbits will not change our EMRI rate by more than a factor of 2.


\vspace{20pt}

\section{Results}
\label{sec:results}

We run four sets of simulations of $1000$ BHs each embedded in a NSC composed of other BHs and $1$~M$_\odot$ stars, as described in Sections \ref{sec: semianalytic_model} and \ref{sec:processes}. We list the simulations and their initial conditions in Table \ref{tab:bh_summary}. We discuss the results and implications below.

\begin{table*}[t]
\centering
\caption{A summary of black hole growth and merger properties across our simulations. We use stellar profile $\alpha=1.25$ for all the following simulations. The merger column tallies the total number of BH-BH captures that occur over the simulation, with the exception of H18*, which refers to the simulation with H18 initial conditions that only considers stellar collisions with no GW capture. We note that some of these captures can occur to the same BH, leading to multiple generations of mergers. We also show the maximum BH mass and spin achieved in each of our simulations, as well as the number of BHs that double in mass, grow by an order of magnitude, and grow by two orders of magnitude. Most notably, the last two simulations with GW capture produced IMBHs, as well as the stellar collisions-only simulation. We note that $\sim$4\% of BHs were initially IMBHs ($> 100$~M$_\odot$) in the H18\raisebox{0.3ex}{\footnotesize{${+}$}}M case. 
}
\renewcommand{\arraystretch}{1.1} 
\footnotesize
\begin{tabular}{l c c c c c c}
\hline
\textbf{IC} & \textbf{Mergers} & \textbf{Max Mass BH $[M_{\odot}]$} & \textbf{Max Spin BH} & \textbf{BHs $>2\times M_i$} & \textbf{BHs $>10\times M_i$} & \textbf{\% BHs $>100~ M_\odot$}\\
\hline
K20 & 34  & 28.4  & 0.712 & 8 & 0 & 0\\
K20 \raisebox{0.3ex}{\footnotesize{$+$}}M  & 30  & 57.8  & 0.844 & 12 & 0 & 0\\
H18  & 371 & 407.3 & 0.805 & 79 & 4 & 7.8\%\\
H18\raisebox{0.3ex}{\footnotesize{$+$}}M  & 535 & 526.0 & 0.855 & 103 & 8 & 14.3\%\\
H18*  &  N/A & 123.7 & 0.196 & 0 & 0 & 2.3\%\\
\hline
\end{tabular}
\label{tab:bh_summary}
\end{table*}

\subsection{Effect of Sequential Collisions with Stars on Black Hole Spin} \label{sec:BHstar_coll_spin_effects}


\begin{figure*}
    \centering
    \includegraphics[width=1.005\linewidth]{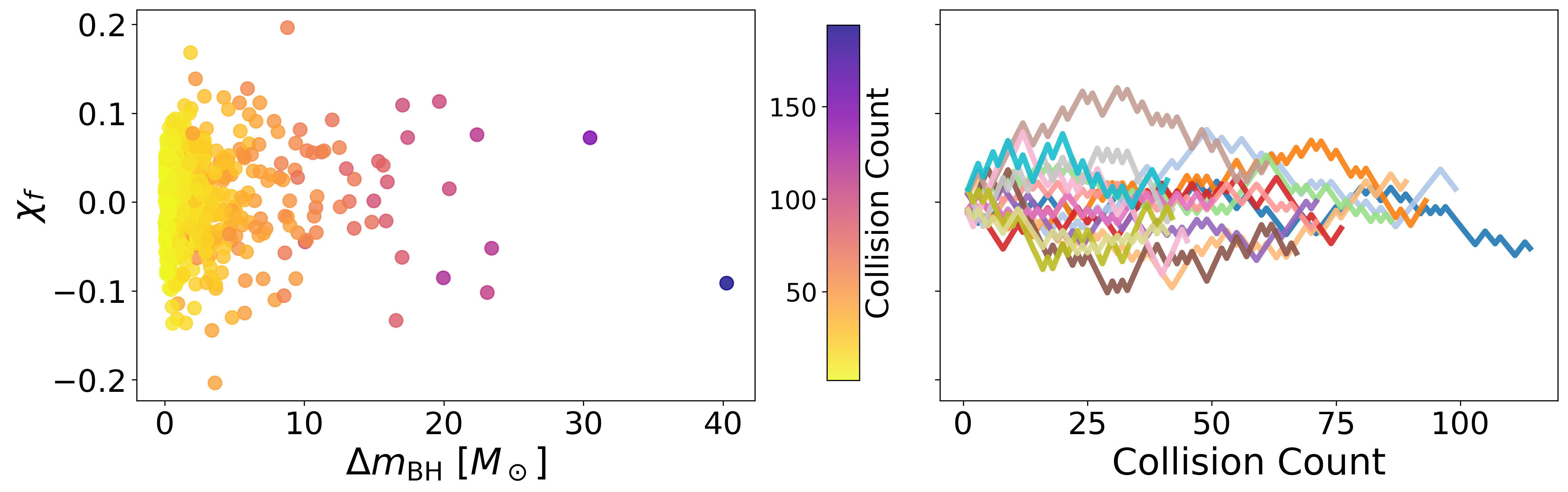}
    \caption{We show the resulting spins and the change in mass of 1000 BHs after 1 billion years with the H18 initial conditions and including only stellar collisions, as described in Section \ref{sec:BHstar_coll_spin_effects} (H18$^\star$) in Table \ref{tab:bh_summary}). \textbf{Left:} The scatter plot shows the final spin $\chi_f$ against the change in final mass from initial mass $\Delta m_{\text{BH}}$. All BHs underwent at least two collisions with stars, with a maximum number of collisions per BH of 194. As shown in Table \ref{tab:bh_summary}, after 10 billion years, 2.3\% of the population are IMBHs with mass above $\sim$ 100 $ M_{\odot}$. \textbf{Right:} We select 16 BHs' sample paths of $\chi$ after each collision for the course of the simulation, demonstrating the random walk behavior and decreasing magnitude of $\chi$ as the BH undergoes successive collisions.}
    \label{fig:final_spins}
\end{figure*}

We begin by isolating the effect of successive BH-star collisions on the BHs by running a simulation without any GW capture. We use the H18 case and show the results of this simulation in the last row of Table \ref{tab:bh_summary}. We evolve BH spin from collisions with stars using the equations described in Section \ref{sec:stellcoll_spin_methodology} to account for changes in a BH's spin due to the accretion of stellar material \citep[e.g.,][]{Gammie2004,Volonteri2005,Volonteri2013}. In the case of multiple accretion episodes, BHs ``spin down'' when the accretion disk rotates in the opposite direction of the BH's initial spin, $\chi_i$, while BHs ``spin up" if they accrete mass from a disk rotating in the same direction as spin $\chi_i$. A larger $\Delta m_{\text{BH}}$, or mass accreted, corresponds to a larger change in spin, $\Delta \chi$ (see Eq. \eqref{eq:coll_af}). However, for the same $\Delta m_{\text{BH}}$, retrograde accretion leads to a greater magnitude of spin change. When the accretion disk is retrograde with respect to the BH spin, the ISCO is further from the BH and therefore carries more angular momentum than when the accretion disk is prograde. Angular momentum conservation demands a greater change in the BH's spin upon accreting the material. Therefore, for the same $\Delta m_{\text{BH}}$, retrograde accretion leads to a greater magnitude spin change. 

With each collision, we randomly draw whether the accretion disk is prograde or retrograde with respect to the BH's initial spin, treating this process as a random walk. We show the final spin parameter versus total change in mass from this simulation in Figure~\ref{fig:final_spins}. The BHs which experienced the most mass growth from successive collisions have lower spins. Our results are consistent with previous findings related to retrograde accretion \citep{Miller2002,
Hughes2003, Volonteri2005, Mandel2007, Volonteri2013,Borchers+2025, Kiroglu+25_highspinsaccretion}; higher mass BHs have lower spins after many successive mergers. Our simulation achieves this same result for successive collisions in galactic nuclei. We therefore expect IMBHs formed through successive BH-star collisions to be low-spinning, with $\chi<0.2$.

\vspace{40pt}

\subsection{Retention of Merger Products} \label{sec:retention}

The magnitude of a GW recoil kick depends in part on the mass ratio of the merging BHs and their spins. While we assume all BHs are initially non-spinning, stellar collisions can modify the spins of these BHs. We find that $\sim $92\% of single BHs have non-zero spins prior to a GW merger due to collisions with stars
; however, the magnitude of $\Delta\chi$ from stellar collisions is low, as shown in Figure \ref{fig:final_spins}. 
We can understand where GW recoil kicks become important by comparing its characteristic speed to the orbital speed as a function of distance from the SBMH. For low-spinning BHs ($\chi < 0.1$), typical kick speeds are between $0$ and $\sim 200$~$\mathrm{km}/\mathrm{s}$ for a range of mass ratios. The Keplerian speed exceeds $\sim 100$ km/s within about $0.03$~pc. As expected, all of our merger products are retained in this steep potential. In fact, none of recoil kicks changed the semi-major axis of the BH's orbit by more than an order of magnitude. However, we note that the orbital eccentricity can increase the recoil speed. \citet{Sopuerta+07} found this increase to be a factor of $1+e$, while numerical relativity studies indicate that an eccentric orbit can boost the recoil speed as much as $25 \%$ \citep{Sperhake+20,Radia+21}. GW bremsstrahlung produces eccentric systems \citep[e.g.,][]{Rom+24_BHcaptureinAGN}. In the most eccentric case, the kick speed could be a factor of $2$ larger. While NSCs should still be able to retain a large fraction of merger products, we reserve a more complete exploration of the effect of BH birth spins and orbital eccentricity on the retention fraction for future work.

\vspace{40pt}



\begin{figure*}
    \centering
    \includegraphics[width=0.8\linewidth]{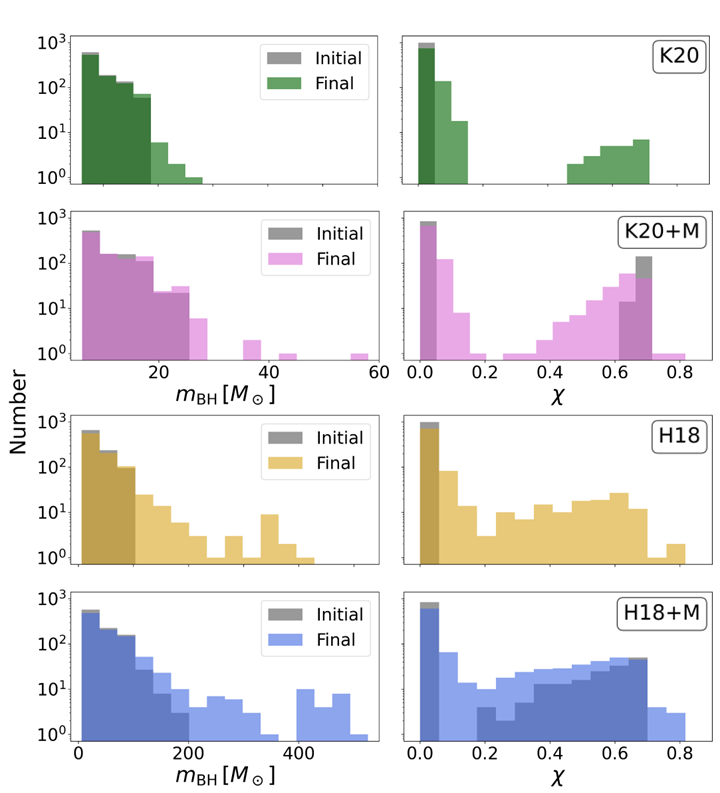}
    \caption{\textbf{Left:} We show the distributions of initial mass in gray and final mass in color for the four initial conditions described in Section \ref{sec:initialconditions}. Populations with primordial binary mergers (K20 \raisebox{0.3ex}{\footnotesize{$+$}}M and H18\raisebox{0.2ex}{\footnotesize{$+$}}M) and the upper mass limit (H18 and H18\raisebox{0.3ex}{\footnotesize{$+$}}M) have correspondingly higher final masses. \textbf{Right:} We plot the distributions of initial spin in gray and final spin in color for the four initial conditions described in Section \ref{sec:initialconditions}. K20  and K20 \raisebox{0.3ex}{\footnotesize{$+$}}M have higher peaks at $\sim 0.7$, and H18 and H18\raisebox{0.3ex}{\footnotesize{$+$}}M have more continuous distributions.}
    \label{fig:final_mass_and_spins_dist}
\end{figure*}

\begin{figure*}
    \centering
    \includegraphics[width=0.78\linewidth]{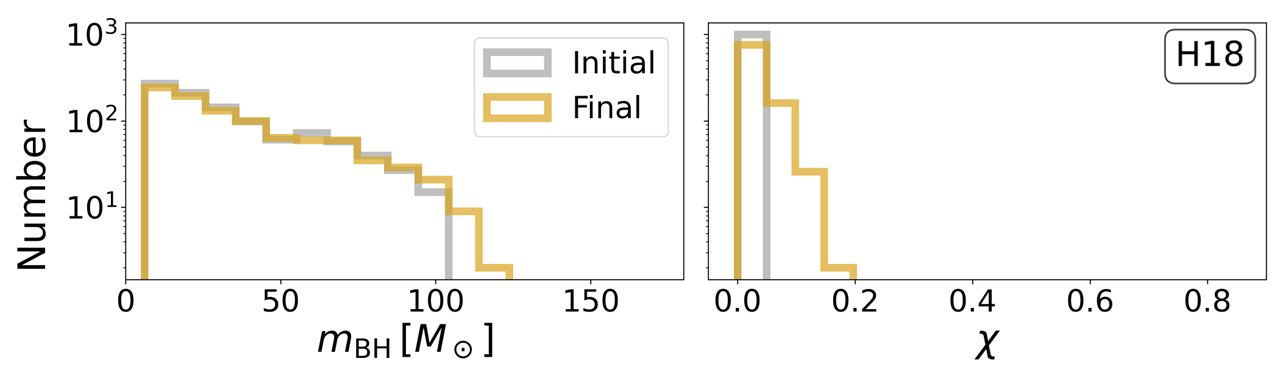}
    \caption{Similar to Figure \ref{fig:final_mass_and_spins_dist}, we show the initial and final distributions of the H18 simulation with only BH-star collisions and stellar profile $\alpha=1.25$. We see less mass growth and spin change than for the simulations in Figure \ref{fig:final_mass_and_spins_dist}.}
    \label{fig:stellcoll_mass_and_spin_dists}
\end{figure*}

\begin{figure}
    \centering
    \includegraphics[width=1.\linewidth]{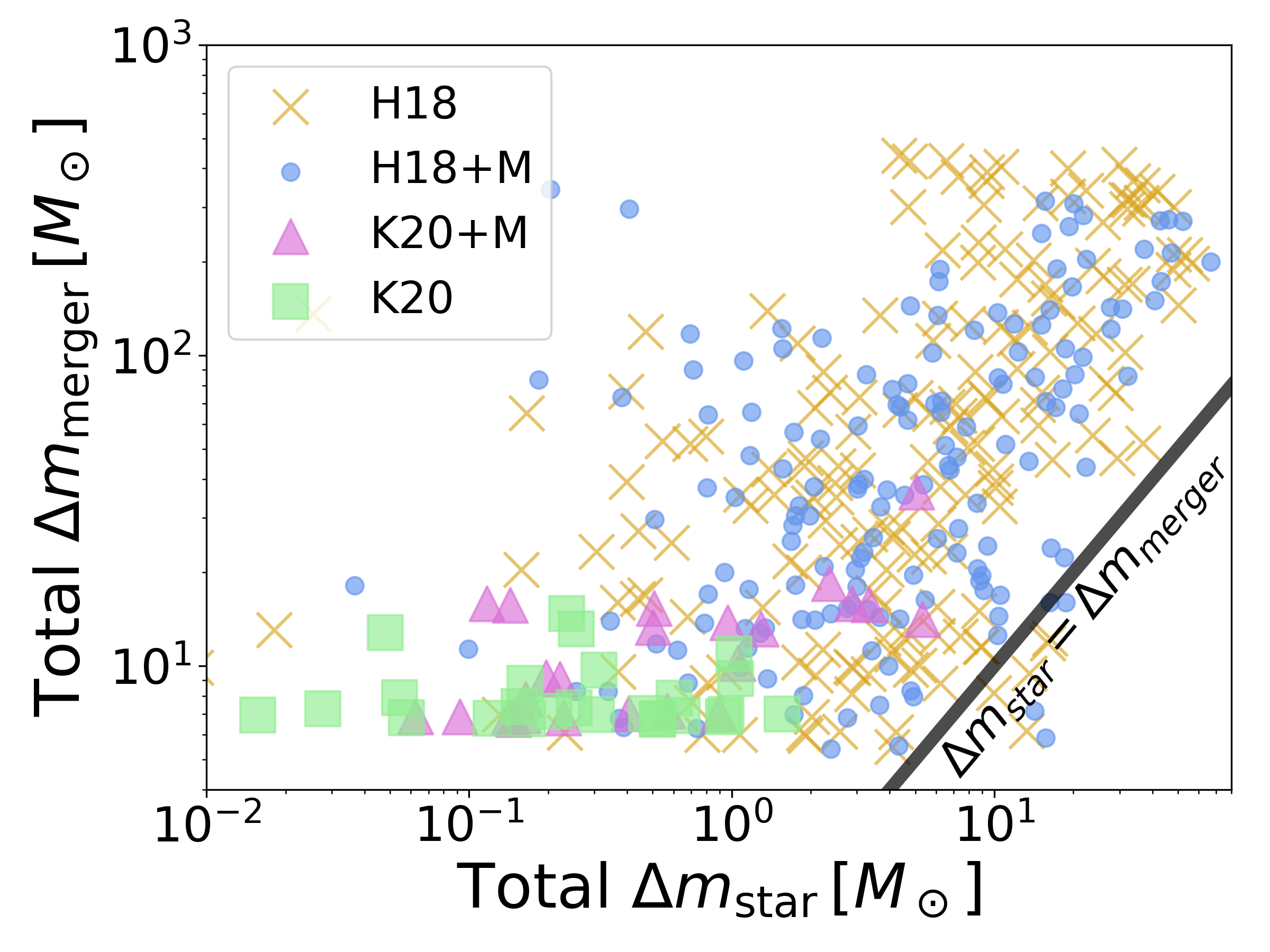}
    \caption{Contribution of each dynamical channel to that mass growth for each BH with $m_{f}>m_{i}$. The x-axis shows the amount of mass accreted over the entirety of the simulation from direct collisions with stars, which is limited by both feedback during the accretion process and environmental conditions like stellar density and velocity dispersion. The y-axis shows the growth through BH-BH GW capture. For all four sets of simulations, most mass growth for the BHs came from GW capture. To guide the eye, we plot a line in black showing where the change in mass from BH-BH mergers and BH-star collisions are equal.}
    \label{fig:dM_star_v_merger}
\end{figure}


\subsection{Comparing Initial and Final Mass and Spin Distributions} \label{sec:mass_and_spin_distributions}

Figure~\ref{fig:final_mass_and_spins_dist} shows the final vs. initial BH mass and spin distributions from our simulations. In order to understand the mass growth of the BHs, we can compare the maximum final mass to the maximum initial mass from the distributions. 
For initial conditions with primordial binaries, the initial maximum BH mass is roughly a factor of $2$ higher than all initially single BHs. Primordial binaries therefore lead to more mass growth for the BH population overall. As can be seen in Table~\ref{tab:bh_summary} and Figure~\ref{fig:final_mass_and_spins_dist}, more BHs double in mass over the course of the simulation for initial conditions with previously merged BHs from primordial binaries. However, the maximum mass has the same order of magnitude. Both lower limit cases K20  and K20 \raisebox{0.3ex}{\footnotesize{$+$}}M resulted in similar scales of mass growth for the BHs; the same is true for both upper limit distributions H18 and H18\raisebox{0.3ex}{\footnotesize{$+$}}M. K20 leads to a maximum final BH mass of $28.4$~M$_\odot$, compared to $57.8$~M$_\odot$ with K20\raisebox{0.3ex}{\footnotesize{$+$}}M. H18 and H18\raisebox{0.3ex}{\footnotesize{$+$}}M produce maximum masses of $407.3$~M$_\odot$ and $526.0$~M$_\odot$, respectively. 

For each BH in our sample, we show its change in mass from capture of other BHs versus direct collisions with stars in Figure~\ref{fig:dM_star_v_merger}. The mass growth of the sample BHs comes primarily from GW capture of other BHs, as opposed to accretion from stars during direct collisions. Even though BH-star collisions are more frequent, the mass accreted from these interactions is kept low due to our prescription to estimate mass loss through feedback. We attribute the dependence of the mass growth and merger rate of the BHs on the initial BH mass distribution to the fact that the BH-BH GW capture timescale is roughly proportional to $1/m_{\rm BH}^2$.


Without primordial binary mergers, all BHs are initially non-spinning, while the population with primordial binaries includes BHs with spins less than $\chi = 0.7$. However, the final spin distributions for all cases show a more continuous distribution. For the K20  and K20 \raisebox{0.3ex}{\footnotesize{$+$}}M cases, which experienced less GW mergers and accrete less mass from stars, there is a sharper peak at about 0.7. Contrastingly, for the H18 and H18\raisebox{0.3ex}{\footnotesize{$+$}}M cases that experience more mass growth through mergers and collisions with stars, the distribution appears almost uniform between $\chi = 0.25$ and $0.7$, with a less sharp peak at $0.7$. All distributions exhibit a spread of low-spinning BHs within $ \chi \lesssim 0.2$, a signature of accretion of material during many successive collisions with stars. 

To highlight the difference between stellar collisions-only and GW mergers in mass and spin distributions, in Figure \ref{fig:stellcoll_mass_and_spin_dists} we plot a simulation using H18 initial conditions and including only collisions with stars on a $\alpha=1.25$ density profile. Figure \ref{fig:dM_star_v_merger} suggests more mass on average is gained through mergers, and this effect is visible in the larger masses produced in Figure \ref{fig:final_mass_and_spins_dist} than Figure \ref{fig:stellcoll_mass_and_spin_dists}, as well. The final mass distribution shows mass accretion through stars, but significantly less overall mass growth than the other simulations with GW mergers. Additionally, the spin peak at 0.7, a hallmark of GW mergers, is not present for the stellar collision-only simulation, whereas the diminishing spin over successive collisions, as in Figure \ref{fig:final_spins}, is present. As shown in the last row of Table \ref{tab:bh_summary}, IMBHs represent $\sim 2 \%$ of the population, but this is largely the result of the high initial mass distribution.

\begin{figure*}
    \centering
    \includegraphics[width=1\linewidth]{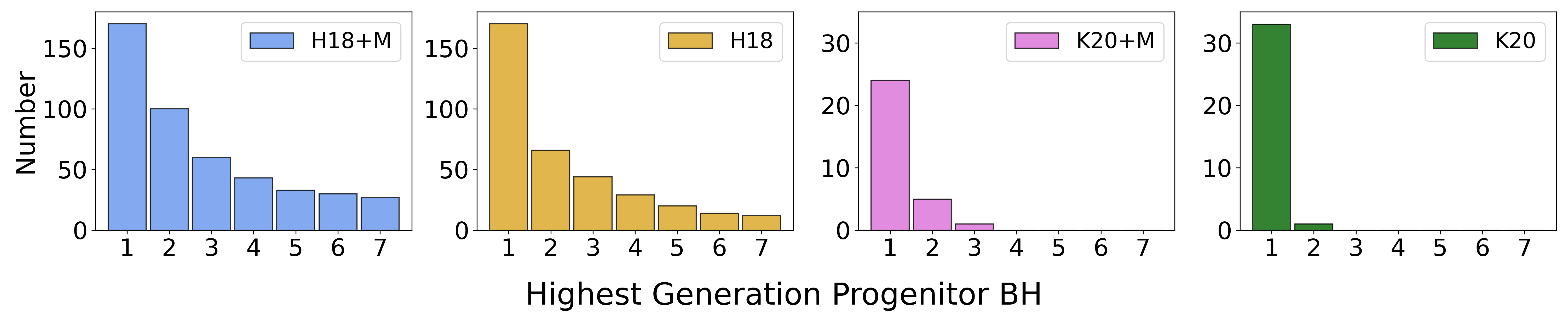}
    \caption{For four simulations described in Table \ref{tab:bh_summary}, number of mergers from GW capture per highest generation BH progenitor in the event. Only GW events from GW capture are included in the number of mergers; events from primordial binary mergers are not included. However, in calculating the generation of the BH, we account for both previous GW captures in the simulation and primordial mergers from the initial conditions. We find that the presence of primordial binaries increases the merger rate with 2G BHs  by approximately a factor of 2 in both populations. The initial masses of the BHs appear to have the strongest effect on the number of mergers with higher generation progenitor BHs. We only show BH generations up to 7G, but the shape can be extrapolated to 7G+ BHs. The highest generation BH progenitor was 16G (12G) for H18\raisebox{0.3ex}{\footnotesize{$+$}}M (H18).
    }
    \label{fig:mergers_gens}
\end{figure*}

\begin{figure*}
    \centering
    \includegraphics[width=0.85\linewidth]{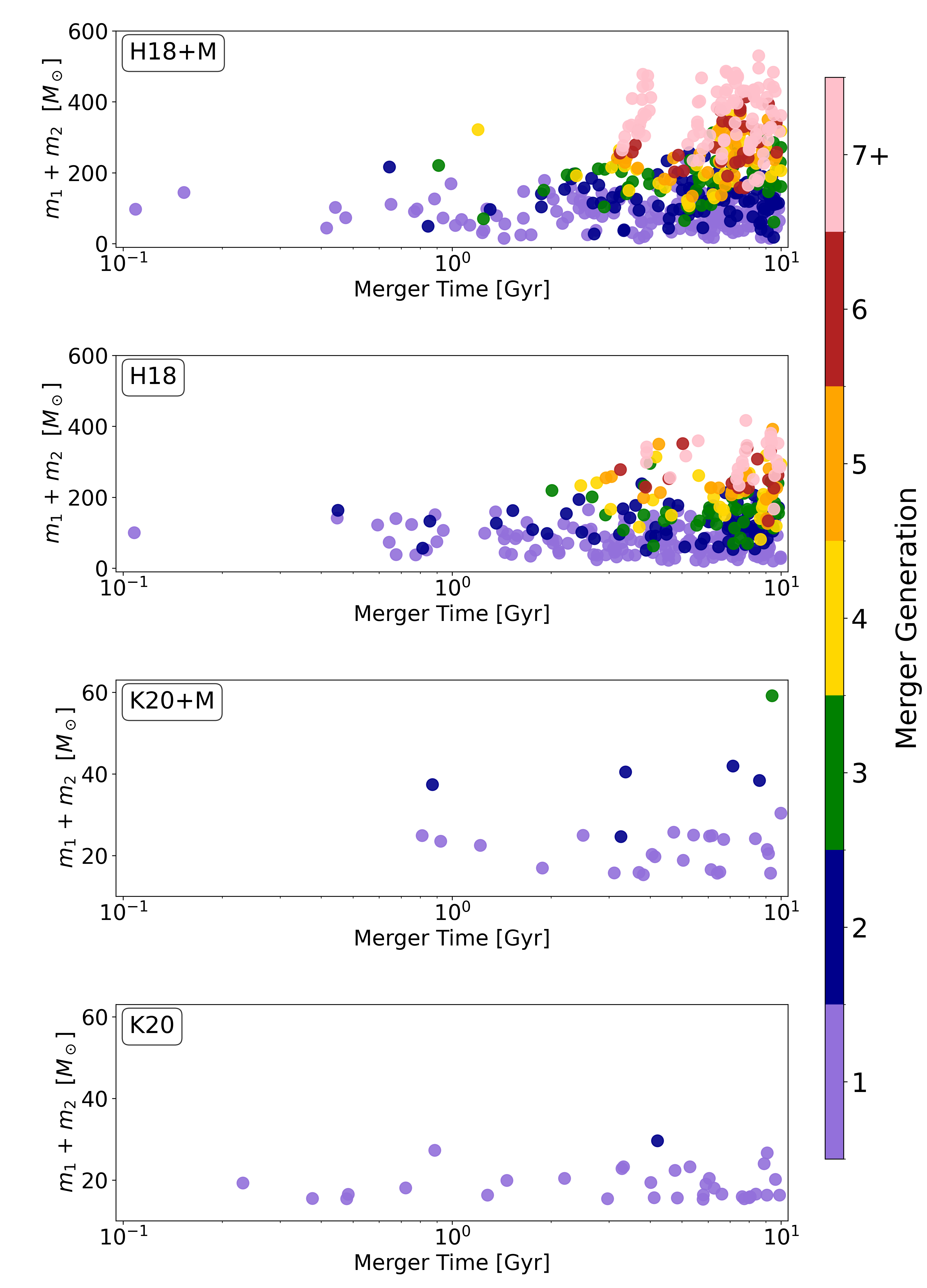}
    \caption{For the four simulations described in Table \ref{tab:bh_summary}, the sum of the mass of the two merging BHs is shown against the time of the merger. Consistent with Figure \ref{fig:mergers_gens}, H18 and H18\raisebox{0.3ex}{\footnotesize{$+$}}M have mergers with higher generation BHs than the H18 and H18\raisebox{0.3ex}{\footnotesize{$+$}}M populations. Across all populations, mergers with higher generation BHs take place later ($\gtrsim1$ Gyr) in the simulation.}
    \label{fig:merger_mass_vs_time}
\end{figure*}

\begin{figure*}
    \centering
    \includegraphics[width=0.85\linewidth]{ 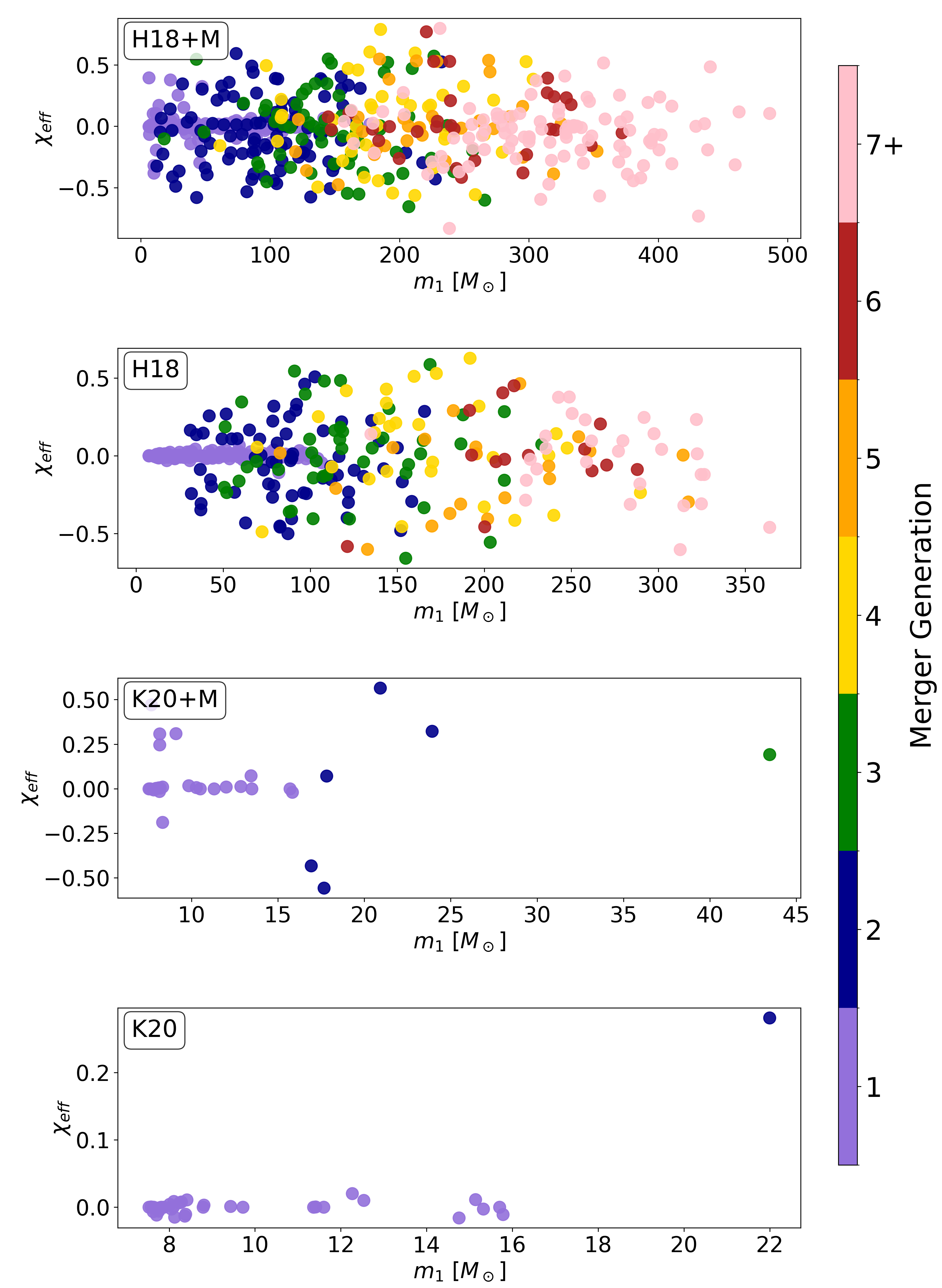}
    \caption{For the four simulations described in Table \ref{tab:bh_summary}, the resulting $\chi_{\rm eff}$ from a merger event is plotted against the initial mass of $m_1$. For all populations except K20, the range of $\chi_{\rm eff}$ is approximately -0.5 to 0.5. As $m_1$ increases, $\chi_{\rm eff}$ increases as well, whereas the most massive $m_1$ in a population have $\chi_{\rm eff}$ with less magnitude, consistent with Figure \ref{fig:final_spins}. It is noted that any non-zero spins of first generation BHs is due to accretion from stellar collisions. All populations had a high percentage ($\sim$ 92\%) of merging BHs with initially non-zero spin due to stellar accretion. For K20, the 1G BHs had spins approximately zero.}
    \label{fig:chieff_vs_mass}
\end{figure*}

\begin{figure*}
    \centering
    \includegraphics[width=1\linewidth]{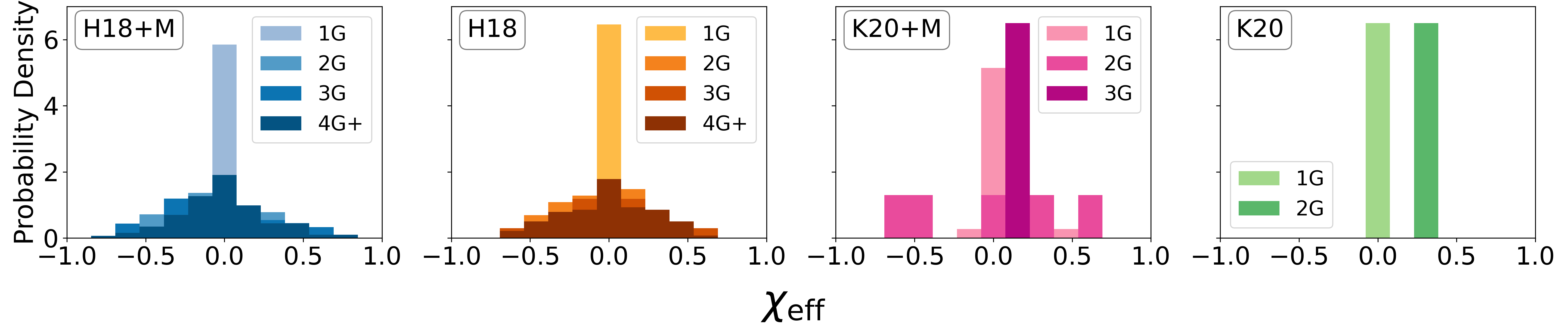}
    \caption{For the four simulations in Table \ref{tab:bh_summary}, the probability density of $\chi_{\rm eff}$ resulting from a merger from GW capture by highest generation BH progenitor. We see that for all populations, mergers with 1G BHs result in low $\chi_{\rm eff}$. In general, mergers with higher generation BHs have larger magnitude $\chi_{\rm eff}$, and the H18 and H18\raisebox{0.3ex}{\footnotesize{$+$}}M populations have larger $\chi_{\rm eff}$ than K20  and K20 \raisebox{0.3ex}{\footnotesize{$+$}}M.} 
    \label{fig:chi_eff_histogram}
\end{figure*}

\subsection{IMBH Formation} \label{sec:IMBHformation}

We assess whether any BHs in our sample population experience significant enough mass growth to become IMBHs, defined as a BH with mass $> 100$~M$_\odot$. We note that the H18 mass distribution with single BHs extends up to but does not exceed $100$~M$_\odot$. However, with primordial binary mergers, the H18\raisebox{0.3ex}{\footnotesize{$+$}}M initial conditions already contain IMBHs as part of $\sim$4\% of the population by this definition. As apparent in Figure~\ref{fig:final_mass_and_spins_dist}, K20  and K20 \raisebox{0.3ex}{\footnotesize{$+$}}M do not result in any IMBHs with $\alpha = 1.25$ for the stellar cusp. The steepness of the stellar cusp can affect the mass growth of the BHs because it controls the number density of stars with which the BHs can collide. Previously, \citet{Rose+22} showed that BH-star collisions alone can produce $\sim 100$~M$_\odot$ IMBHs with Bahcall-Wolf profile ($\alpha = 1.75$) for the stars. However, for the initial conditions that we sample in this study, only H18 and H18\raisebox{0.3ex}{\footnotesize{$+$}}M consistently produce IMBHs. Crucially, our results indicate that these upper limit initial conditions can produce IMBHs from either sequential BH mergers or successive BH-star collisions alone. Specifically, IMBHs represent $14.3 \%$ ($7.8 \%$) of the final population for the H18\raisebox{0.3ex}{\footnotesize{$+$}}M (H18) initial distribution, as shown in Table \ref{tab:bh_summary}. In other words, both channels working in concert with one another are not necessary to produce IMBHs. 
We note that IMBHs grown solely through successive BH–star collisions tend to end up with lower spins than those formed through BH–BH mergers. This trend is partly driven by our simplified spin prescription for BH-star collisions, which assumes fully anti-aligned accretion during spin-down and therefore represents a lower limit on the resulting spin distribution.
We reserve the exploration of other initial conditions and cluster properties for future work, such as whether there is a mass distribution between our lower and upper limits that consistently produces IMBHs.

The most massive BH formed across all of our simulations was $526$~M$_\odot$. As a BH become significantly more massive than the rest of the population, it begins to sink towards the SMBH due to dynamical friction until it becomes an EMRI. All BHs with final mass $>400$~M$_\odot$ become EMRIs. Furthermore, for H18 (H18\raisebox{0.3ex}{\footnotesize{$+$}}M) initial conditions, 70\% (55\%) of BHs over $>200$~M$_\odot$ became EMRIs. 
These events still have mass ratio $<10^{-3}$ with the SMBH and therefore do not meet the criterion for intermediate mass ratio inspirals (IMRIs). However, they may be a signature of efficient dynamical formation of IMBHs in galactic nuclei. These heavier EMRIs occur at later times within the cluster, after dynamical formation channels like GW capture have had time to act and produce IMBHs through sequential mergers. The rate of EMRIs with mass ratio $>5 \times 10^{-5}$ is approximately $4$ Gyr$^{-1}$ per Milky Way-like galaxy,
and they predominantly occur after $\sim 2$~Gyr in the simulation. EMRIs with mass ratio $>1 \times 10^{-4}$
predominantly occur after $4$~Gyr in the simulation at a similar rate of $4.8$ Gyr$^{-1}$ per Milky Way-like galaxy. For the range of initial conditions and cluster parameters explored in this study, IMBHs with masses $>10^4$~M$_\odot$ cannot form in situ in a NSC, though they may instead be deposited in the NSC through other means, such as an infalling globular cluster \citep[e.g.,][]{EGP+25}. We note as a caveat on these results that our simulations assume a density profile for the background BHs and stars that are held constant for the entirety of the simulation. We do not account for the effects of an IMBH scattering the relatively lighter BHs and stars to wider orbits as it sinks towards the SMBH \citep[e.g.,][]{Baumgardt+06b,LockmannBaumgardt08}. The formation of an IMBH may temporarily suppress the ability of dynamical channels to form another IMBH until the cusp has been replenished (estimated to be on $\gtrsim 50$~Myr timescales for the stars by aforementioned studies).

\subsection{Implications for GW Sources} \label{sec:GW_rates_and_properties}

In Figure~\ref{fig:mergers_gens}, we show the number of GW events from GW capture per the highest generation progenitor BH. From this, we can estimate a merger rate from GW capture for different BH generations. In populations containing primordial binaries, we classify the remnants of primordial BBH mergers as 2G, consistent with dynamically formed BBHs.

K20 initial conditions give $30$ mergers with 1G BHs over our simulation time, yielding a rate of a few $\times 10^{-9}$ yr$^{-1}$ per Milky Way-like galaxy. This value falls within the range previously computed in the literature \citep{OLeary+09}, where the rate depends on assumptions about the ratio of BHs to stars, BH mass range, and slope of the cusp \citep[][]{OLeary+09,Rose+22,liu2025}. Mergers in which one or both of the progenitor BHs is 2G, the product of a previous merger, occur at a much lower rate of a few $\times 10^{-10}$ yr$^{-1}$ per Milky Way-like galaxy. 

The overall rate of GW capture is comparable between K20  and K20 \raisebox{0.3ex}{\footnotesize{$+$}}M initial conditions. However, with K20\raisebox{0.3ex}{\footnotesize{$+$}}M, mergers that contain a 2G BH represent a higher fraction of the systems. The merger rate for 1G progenitor BHs is $\sim 2 \times 10^{-9}$ yr$^{-1}$ per Milky Way-like galaxy, whereas the 2G progenitor BH rate is  $\sim 5 \times 10^{-10}$ yr$^{-1}$ per Milky Way-like galaxy. Mergers with 3G progenitor BHs occur a  few $\times 10^{-10}$ yr$^{-1}$ per Milky Way-like galaxy.


Similarly, we calculate merger rates from GW capture for H18 initial conditions. This simulation gives $\sim150$ mergers with 1G BHs, corresponding to merger rate of $10^{-8}$ yr$^{-1}$ per Milky Way-like galaxy, which is an order of magnitude higher than the rate for both K20  and K20 \raisebox{0.3ex}{\footnotesize{$+$}}M. Additionally, the merger rate for 2G BHs, $5 \times 10^{-9}$ yr$^{-1}$ per Milky Way-like galaxy, is also higher than the corresponding rate for K20  and K20 \raisebox{0.3ex}{\footnotesize{$+$}}M. The merger rate for 1G BHs in the H18\raisebox{0.2ex}{\footnotesize{$+$}}M case is comparable to the H18 rate. However, the H18\raisebox{0.3ex}{\footnotesize{$+$}}M merger rate for 2G BHs is closer to $10^{-8}$ yr$^{-1}$ per Milky Way-like galaxy. The qualitative effect of primordial binary mergers on the merger rate is similar in both K20\raisebox{0.3ex}{\footnotesize{$+$}}M and H18\raisebox{0.2ex}{\footnotesize{$+$}}M distributions: the H18\raisebox{0.3ex}{\footnotesize{$+$}}M initial conditions resulted in more mergers with 2G+ BHs. These mergers represent about twice the fraction of total mergers compared to the H18 simulation. The rate of mergers with a 3G+ BH for both H18 and H18\raisebox{0.3ex}{\footnotesize{$+$}}Mis at most $\sim 5 \times 10^{-9}$ yr$^{-1}$ per Milky Way-like galaxy, an order of magnitude higher than the corresponding rate for K20 \raisebox{0.3ex}{\footnotesize{$+$}}M. The effect of primordial binaries is significant in increasing the number of 2G BHs in the population, and therefore higher generation BHs, as shown in Figure \ref{fig:mergers_gens}.

For all mergers from GW capture, we show the total mass versus the time of merger in Figure~\ref{fig:merger_mass_vs_time}, color-coded by the highest generation progenitor BH. Generally, the mergers with the largest masses occur at later times. Mergers with higher generation BHs (3G+) occur predominantly in the H18 and H18\raisebox{0.3ex}{\footnotesize{$+$}}M simulations. The mergers with 
more massive and higher generation BHs occur $\sim 2$ Gyr into the simulation. Additionally, higher generation BHs are more likely to occur in any populations with primordial binaries.


To complement Figure~\ref{fig:merger_mass_vs_time}, we also show the $\chi_{\text{eff}}$ of these GW sources in Figures~\ref{fig:chieff_vs_mass} and \ref{fig:chi_eff_histogram}. Specifically, we show the $\chi_{\text{eff}}$ versus primary mass in Figure~\ref{fig:chieff_vs_mass}, color-coded by generation, while Figure \ref{fig:chi_eff_histogram} shows the $\chi_{\text{eff}}$ distribution for each generation.  For mergers with the highest generation BHs (6G+), they occur almost exclusively when the higher generation BH has mass $>250$~M$_\odot$, shown in Figure \ref{fig:chieff_vs_mass}. Correspondingly, BHs with mass $>250$~M$_\odot$ are almost exclusively 6G+. For all populations other than the last row (K20), the magnitude of $\chi_{\text{eff}}$ for successive mergers can be $\gtrsim 0.5$. As seen in Figure \ref{fig:chieff_vs_mass}, the magnitude of $\chi_{\text{eff}}$ tends to decrease with mergers with 6G+ BHs, consistent with the results in Section \ref{sec:BHstar_coll_spin_effects}. Generally 1G BHs have zero initial spin, but considering accretion from stellar collisions, we see some 1G BHs with initially non-zero spin and therefore non-zero, albeit often very small, $\chi_{\text{eff}}$.

\section{Conclusions}

We consider two channels for modifying the mass and spin distributions of BHs in NSCs:
\begin{enumerate}
    \item \textit{BH accretion through collisions with stars:} From timecale arguments alone, this channel is the most promising: a BH residing within $10^{-2}$~pc of the SMBH will experience tens to hundreds of collisions with stars. However, the effect of these sequential collisions on the BHs is limited by the amount the BHs can conceivably accrete from stars during each collision.
    \item \textit{BH mergers:} For a population of single BHs, mergers can occur following GW capture. While the timescale for GW capture is much longer than direct BH-star collisions, this process has the advantage of producing appreciable changes in a BH's mass and spin. 
\end{enumerate}

We assess the interplay of these two channels using the semianalytic model described in Section \ref{sec: semianalytic_model}. We consider four sets of initial conditions. 
In two of these initial conditions, we account for the possibility that a subset of our initial population are the merger products of primordial binaries following \citet{Hoang+18}. The first two initial conditions, K20  and K20 \raisebox{0.3ex}{\footnotesize{$+$}}M, (color-coded as green and pink in Figure~\ref{fig:IC_plot}) adopt lower initial masses for the BHs \citep[for the mass distribution, see][]{Kremer+20}. With primordial binary mergers (K20 raisebox{0.3ex}{\footnotesize{$+$}}M), the lower limit extends up to roughly $\sim 25$~M$_\odot$ and has BHs that are initially spinning, while without primordial binaries (K20), all BHs are initially non-spinning and have masses extending up to $15$~M$_\odot$. We also test an upper limit mass distribution, which extends to $\lesssim 100$~M$_\odot$ without primordial binaries (H18) and $\lesssim 200$~M$_\odot$ with primordial binaries (H18\raisebox{0.3ex}{\footnotesize{$+$}}M), as in \citet{Hoang+18}. We find the following:

\begin{enumerate}

\item \textbf{BH Mass and Spin Distributions:} The lower mass limit (K20  and K20 \raisebox{0.3ex}{\footnotesize{$+$}}M) and upper mass limit (H18 and H18\raisebox{0.3ex}{\footnotesize{$+$}}M) experience similar scales of mass growth, respectively. For the initial conditions with primordial binaries (K20 \raisebox{0.3ex}{\footnotesize{$+$}}M and H18\raisebox{0.2ex}{\footnotesize{$+$}}M), more BHs within the population grow in mass. The most massive final BHs for the K20  and K20 \raisebox{0.3ex}{\footnotesize{$+$}}M initial distributions have the same order of magnitude, while the same is true for the H18 and H18\raisebox{0.3ex}{\footnotesize{$+$}}M simulations. The maximum final BH mass from H18 and H18\raisebox{0.3ex}{\footnotesize{$+$}}M distributions is an order of magnitude larger than the most massive BH from K20  and K20 \raisebox{0.3ex}{\footnotesize{$+$}}M distributions. As can be seen in Figure \ref{fig:final_mass_and_spins_dist}, populations with initially more massive BHs experience more overall mass growth. This is in part because the BH-BH GW capture timescale $\propto \frac{1}{m_{BH}^2}$, while more massive BHs also have a slightly larger cross-section due to gravitational focusing, shortening the collision timescale, and can retain more of a star's mass during a collision. We conclude that the range of the initial mass distribution is a better predictor the final BH masses than the presence of primordial binaries. 

\item \textbf{Prospects for Forming IMBHs:} The maximum BH mass produced through dynamical channels in our simulations was $\sim 500$~M$_\odot$. Dynamical friction limits the mass of BHs formed in situ to hundreds of ~M$_\odot$. As the IMBH becomes significantly more massive than other objects in the cluster, it should start sinking towards the SMBH, eventually becoming an EMRI. IMBH formation also depends on the initial mass distribution: IMBHs are not formed at all in the K20  and K20 \raisebox{0.3ex}{\footnotesize{$+$}}M populations with stellar profile $\alpha = 1.25$.


\item \textbf{Rate of BH-BH Mergers:}
We find the merger rate is a few $\times 10^{-9}$ per year per Milky Way-like galaxy for K20  and K20 \raisebox{0.3ex}{\footnotesize{$+$}}M, assuming a BH density profile from \citet{AharonPerets16}. For H18 and H18\raisebox{0.3ex}{\footnotesize{$+$}}M, we see a merger rate of an order of magnitude higher, a few $\times 10^{-8}$ per year per Milky Way-like galaxy. For mergers with 2G BHs, the rates are $\sim 1 \times 10^{-8}$ (H18\raisebox{0.3ex}{\footnotesize{$+$}}M), $\sim 5 \times 10^{-9}$ (H18), $\sim 5 \times 10^{-10}$ yr$^{-1}$ (K20 \raisebox{0.3ex}{\footnotesize{$+$}}M), $\sim 1 \times 10^{-10}$ (K20) yr$^{-1}$ per Milky Way-like galaxy. These results demonstrate the effect of primordial binaries on the merger rate for the lower mass limit, as the 2G rate for the H18\raisebox{0.3ex}{\footnotesize{$+$}}M is an order of magnitude higher than H18, and the 2G rate for the K20 \raisebox{0.3ex}{\footnotesize{$+$}}M is roughly half an order of magnitude higher than K20. Additionally, K20 \raisebox{0.3ex}{\footnotesize{$+$}}M contains a higher (3G) generation BH in a merger; K20 does not. 

\item \textbf{Other Implications for GW Sources:}
We find that mergers with \textbf{higher generation BH progenitors} (3G+) occur primarily in the H18 and H18\raisebox{0.3ex}{\footnotesize{$+$}}M populations and after $\sim 2$ Gyr. Additionally, the H18\raisebox{0.3ex}{\footnotesize{$+$}}M population with primordial binaries increase the likelihood of mergers with higher generation BHs. The mergers with the highest generation BH progenitors (6G+) have a primary BH with mass $> 250~$M$_\odot$. The expected $\chi_{\text{eff}}$ depends on the generation of BH. For first generation BHs, the $\chi_{\text{eff}}$ is approximately zero, though BHs can have non-zero spin as the result of accretion from stellar collisions. For high generation BHs (3G+), the $\chi_{\text{eff}}$ has magnitude of up to $\sim 0.5$. For the highest generation BHs (6G+), $\chi_{\text{eff}}$ is generally magnitude $< 0.5$ due to the effect of successive stellar collisions and mergers reducing spin, $\chi$, magnitude.


\end{enumerate}

To understand how populations of single BHs in a NSC evolve over time, we model the effects of BH-BH GW capture and BH-star direct collisions semianalytically. The BH-BH merger rates and properties can depend sensitively on assumptions about the BH masses, initial binary fraction, and density profile within the cluster \citep[see also, e.g.,][]{OLeary+09,liu2025}. 
Our work highlights the need to explore the broad range of initial conditions and assumptions about the cluster in the context of GW source production and IMBH formation.

\begin{acknowledgments}
A.N.\ was supported by NSF Grant AST-2149425, a Research Experiences for Undergraduates (REU) grant awarded to CIERA at Northwestern University. S.C.R.\ thanks the CIERA Lindheimer Fellowship for support. F.A.R.\ acknowledges support from NSF Grants AST-2108624 and AST-2511543. F.K.\ acknowledges support from a CIERA Postdoctoral Fellowship.

This research was supported in part through the computational resources and staff contributions provided for the Quest high performance computing facility at Northwestern University which is jointly supported by the Office of the Provost, the Office for Research, and Northwestern University Information Technology.
\end{acknowledgments}

\newpage

\appendix %

\section{Comparison of BH Number Densities}
\label{app:n_BH_comp}
\begin{figure*}[h!]
    \centering
    \includegraphics[width=0.43\linewidth]{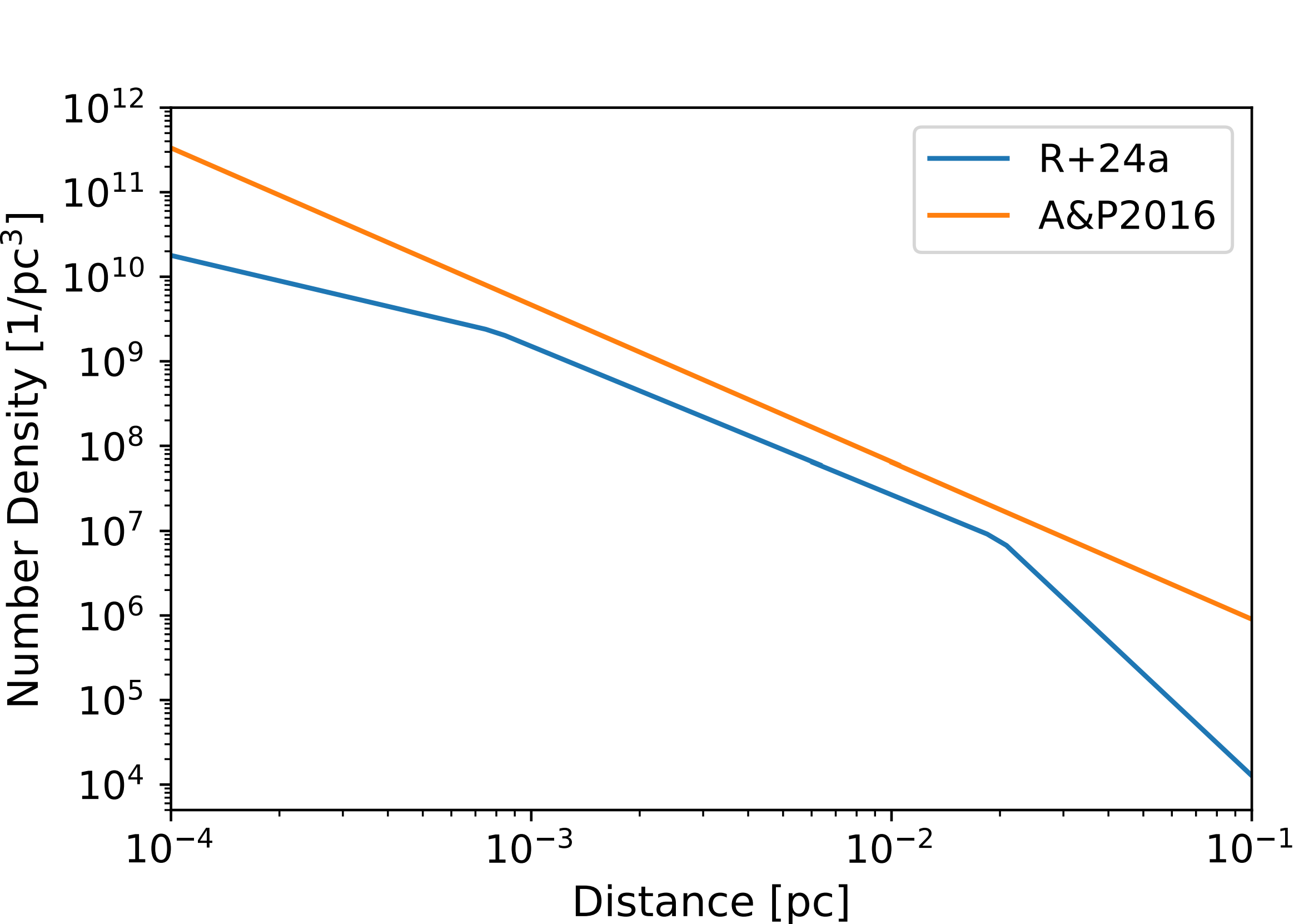}
    \caption{A comparison of the number densities of BHs using a power law fit from \citet{AharonPerets16} (A\&P2016) and a piecewise function from \citet{Rom+2024} (R+24a).}
    \label{fig:density_comp}
\end{figure*}

We plot the number densities of BH from \citet{AharonPerets16} and \citet{Rom+2024} in Figure \ref{fig:density_comp}. We note that the latter presents a four-part piecewise function, while the former offers a density profile well-represented by a power-law. The two densities are within an order of magnitude of each other from about $10^{-3}$ to $10^{-2}$ pc, but differ up to three orders of magnitude by $10^{-1}$ pc. Since the dynamical processes have long timescales outside 0.1 pc (e.g., $t_{\text{GW}}>>10$ Gyr), we opt to use the simpler power-law fit. We further note that the ratio of the number of BHs to stars is a free parameter in \citet{Rom+2024}, and they consider two values in their paper, $10^{-4}$ and $10^{-3}$. We use $f_{\text{BH}}$ = $10^{-4}$ in this plot, which produces a break in their density profile around 0.02 pc, where it begins to decrease more steeply. $f_{\text{BH}}$ = $10^{-3}$  would make the break in the density profile occur further from the SMBH, meaning the overall density profile would be more similar to \citet{AharonPerets16}.

\newpage

\bibliography{sample631}{}
\bibliographystyle{aasjournal}

\end{document}